# FORMATION AND EVOLUTION OF CORONAL HOLES DURING THE RISING PHASE OF CYCLE 23

## Bilenko I. A.[1], Tavastsherna K. S.[2]

*1Moscow State University, Sternberg Astronomical Institute, Moscow, Russia,*
e-mail: bilenko@sai.msu.ru
*2Central Astronomical Observatory at Pulkovo of RAS, St. Petersburg, Russia*

Regularities of formation of coronal holes (CH) at the rising phase of cycle 23 are investigated. The period from 01.01.1997 to 01.03.2000 (Carrington rotations (CRs) 1918-2059) is considered in detail. The evolution of the global magnetic field (GMF) of the Sun from the zonal to the sectorial structure is analyzed. It is shown that the zonal structure is quasi-stable. The sum of zonal harmonics dominated up to CR 1941, although a stable four-sector structure of GMF was formed in 1932. In CRs 1941-1950 the contribution of the zonal and sectorial components becomes approximately equal and from CR 1950 the sectorial structure of GMF was dominated. Sectorial structure of GMF undergoing sharp changes from four-sector, at the beginning of growth of the sectorial harmonics (CR 1926), to the two-sector structure, then back to four-sector and then again to the two-sector structure. CHs clearly trace all evolutionary changes in the GMF. The structure of the polarity of the GMF uniquely determines the zones of photospheric magnetic fields where the CHs are formed.

## 1. INTRODUCTION

Coronal holes (CHs) are one of the most important factors determining the space weather and affecting the geomagnetic activity. The origin of both high- and low-speed solar wind streams is related to CHs (Nolte et al., 1976; Sheeley and Harvey, 1981; Temmer et al., 2007). CHs are observed at different levels in the solar atmosphere. The corona images in the EUV and X-ray show them as regions with reduced brightness and according to observations in the HeI 10830 Å chromospheric line, they appear as enhanced brightness objects (Harvey and Sheeley, 1979; Kahler et al., 1983).

In several studies it was found that CHs are located in the unipolar magnetic field regions (Harvey et al., 1982; Timothy et al., 1975; Varsic et al. 1999), although the degree of unipolarity in the different parts of photospheric magnetic fields associated with the CH region may be different (Bilenko, 2005).

The dynamics of CHs is known to show 11-year and 22-year periodicity associated with the solar activity cycles. There are polar and non-polar CHs. Their dynamics in the cycles of solar activity is very different (Bilenko and Tavastsherna, 2016). Ikhsanov and Ivanov (1999) showed that long-lived equatorial CHs have approximately the same differential rotation as sunspots, while long-lived polar CHs have rigid rotation. In general, CHs well trace evolutionary changes in the global magnetic field (GMF) (Insley et al., 1995; Sanchez-Ibarra and Barraza-Paredes, 1992; McIntosh, 2003; Bilenko and Tavastsherna, 2016). Ivanov and Obridko (2014) noted that large-scale magnetic fields play an important role in the organization of all solar activity. Bilenko and Tavastsherna (2017) have found that the majority of CHs are formed during the periods of stable structure of the GMF. As the rate of structural changes in GMF increases, the number of CHs and the values of their parameters decrease. The correlation between area, extension in latitude and longitude, magnet-flux of CHs and intensity of the GMF is higher in the cycles with higher values of the magnetic field and a more stable structure of the GMF (Bilenko and Tavastsherna, 2017).

Despite the great attention paid recently to the study of CHs, the regularities of their formation, the processes that determine the dynamics of CHs in solar cycles, the relationship of CHs with other phenomena of solar activity are still not clear. Therefore, the study of the regularities of formation of CHs and their evolutionary changes in the solar cycles is important for understanding the processes occurring in the Sun, and for the space weather and geomagnetic activity forecast.

The purpose of this work is a detailed consideration of the formation and dynamics of CHs at the rising phase of cycle 23, as it is at the growth phase of solar activity is a sharp change in the parameters of all processes in the Sun, which has a direct impact on the interplanetary medium and the Earth's magnetosphere.

## 2. DATA

The period from 01.01 1997 to 01.03.2000, which corresponds to Carrington rotations (CRs) 1918-2059 is investigated. Data from the Kitt Peak observatory were used to compare the CHs with photospheric magnetic fields. The CH boundaries obtained from daily observations in the HeI 10830 Å line were superimposed on the data of daily observations of the photospheric magnetic fields of the full disk of the Sun in the line FeI λ=8688 Å, and the synoptic data of the CH locations were superimposed on synoptic maps of magnetic fields.

In the study of the dependences of CH dynamics on GMF dynamics, the synoptic magnetic field maps calculated in a potential approximation on the source surface (2.5 Rs) of the WSO (Wilcox Solar Observatory) (Altschuler and Newkirk, 1969; Altschuler et al. 1975; Hoeksema and Scherrer, 1986; Schatten et al., 1969) were used. The data of harmonic coefficients of the large-scale magnetic field decomposition on the spherical components of the WSO were also used.

Data from the SOHO/EIT space observatory in the FeXII 195 Å line were used to illustrate the location of CHs in the corona.

## 3. EVOLUTIONARY CHANGES OF THE GLOBAL MAGNETIC FIELD DURING THE RISING PHASE OF CYCLE 23

The dynamics of the solar GMF is determined by the evolutionary changes in the large-scale magnetic fields of the Sun, which also govern the interplanetary magnetic field, in the behavior of the low-frequency components of the spherical decomposition of the photospheric magnetic fields of the Sun (Berdichevskaya, 1987), in the change of the sign of the magnetic field at the poles in the maxima of solar activity (Harvey and Recely, 2002; Bilenko, 2002). Figure 1a shows the changes in the magnetic fields of positive and negative polarity and the sum of their module obtained from the longitudinal distribution diagram of the magnetic fields of the Sun on the source surface (2.5 Rs), calculated in the potential approximation for each CR for the period under consideration (Bilenko, 2014; Bilenko and Tavastsherna, 2016). The thin lines correspond to the data for each CR, and the thick lines correspond to that averaged for 7 CRs.

The solar magnetic field can be described as a function of latitude and longitude (r, θ, φ) (Altschuler and Newkirk, 1969; Chapman and Bartels, 1940):

$$\psi(r,\theta,\varphi) = R \sum_{n=1}^{N} \sum_{m=0}^{n} \left(\frac{R}{r}\right)^{n+1} [g_n^m \cos(m\phi) + h_n^m \sin(m\phi)] P_n^m(\theta),$$

where: $P_n^m(\theta)$ – are the associated Legendre polynomials, and $N$ is the number of harmonics. The coefficients: $g_n^m, h_n^m$ are calculated by fitting with the least squares method of the observed radial component of photospheric magnetic fields in a potential approximation. On the basis of the coefficients $g_n^m, h_n^m$ it is possible to obtain the power spectrum of various harmonic components from the spherical harmonic decomposition of the global magnetic field of the Sun (Altschuler et al., 1977; Levine, 1977; Stix, 1977):

$$S_n = \sum_{m=0}^{n} [(g_n^m)^2 + (h_n^m)^2].$$

The transition from the zonal structure of the distribution of magnetic fields to the sectorial in the rising phase and the inverse transformation from the sectorial structure of the distribution of magnetic fields to the zonal one in the late declining phase of the of solar activity is a manifestation of the cyclic evolution of GMF. This topological change of magnetic fields is essential, because during this period the general level of solar activity, as well as the parameters of all solar activity phenomena, changes dramatically. Figure 1b shows the changes in the sum of the zonal (dashed line) and the sum of the sectorial harmonics (solid line). The ratio of their values is shown in Figure 1c. Changes in the axisymmetric dipole component (dashed line) and the equatorial dipole (solid line) are shown in Figure 1d. The axisymmetric and antisymmetric relative to the equator harmonics spectrum is shown in Figure 1e, and axisymmetric and symmetric relative to the equator harmonic spectrum is shown in Figure 1f. The evolution of the polar angle of the dipole component is shown in Figure 1g. From Figure 1 it follows that during the rising phase of the cycle 23 with the growth of the magnetic field the contribution of sectorial GMF structures increased, and zonal decreased. The increase in the sectorial components began from about CR 1926, but prior to CR 1940 a zonal structure of the GMF was dominated (Figure 1c). In 1941, the level of the sum of zonal and sectorial harmonics became equal in magnitude, and, since 1950, the sectorial structure of GMF were dominated. The axisymmetric dipole component (dashed line) gradually decreased throughout the rising phase of cycle 23, and the component of the equatorial dipole (solid line) began to grow slowly from CR 1927, approximately simultaneously with the growth of the sectorial component and the growth of the polar angle of the dipole component, and its growth increased sharply from CR 1941, i.e. from the moment when the contribution of sectorial and zone harmonics became equal. The contribution of the axisymmetric and antisymmetric relative to the equator harmonics (Figure 1e) decreased rapidly from CR 1948, and axisymmetric and symmetric relative to the equator harmonic increased (Figure 1f), reaching respectively the maximum and minimum in CR 1955. Then, the level of axisymmetric and antisymmetric harmonics with respect to the equator began to grow, and axisymmetric and symmetrical relative to the equator harmonics declined. The polar angle of the dipole component underwent sharp fluctuations during CRs 1949-1959.

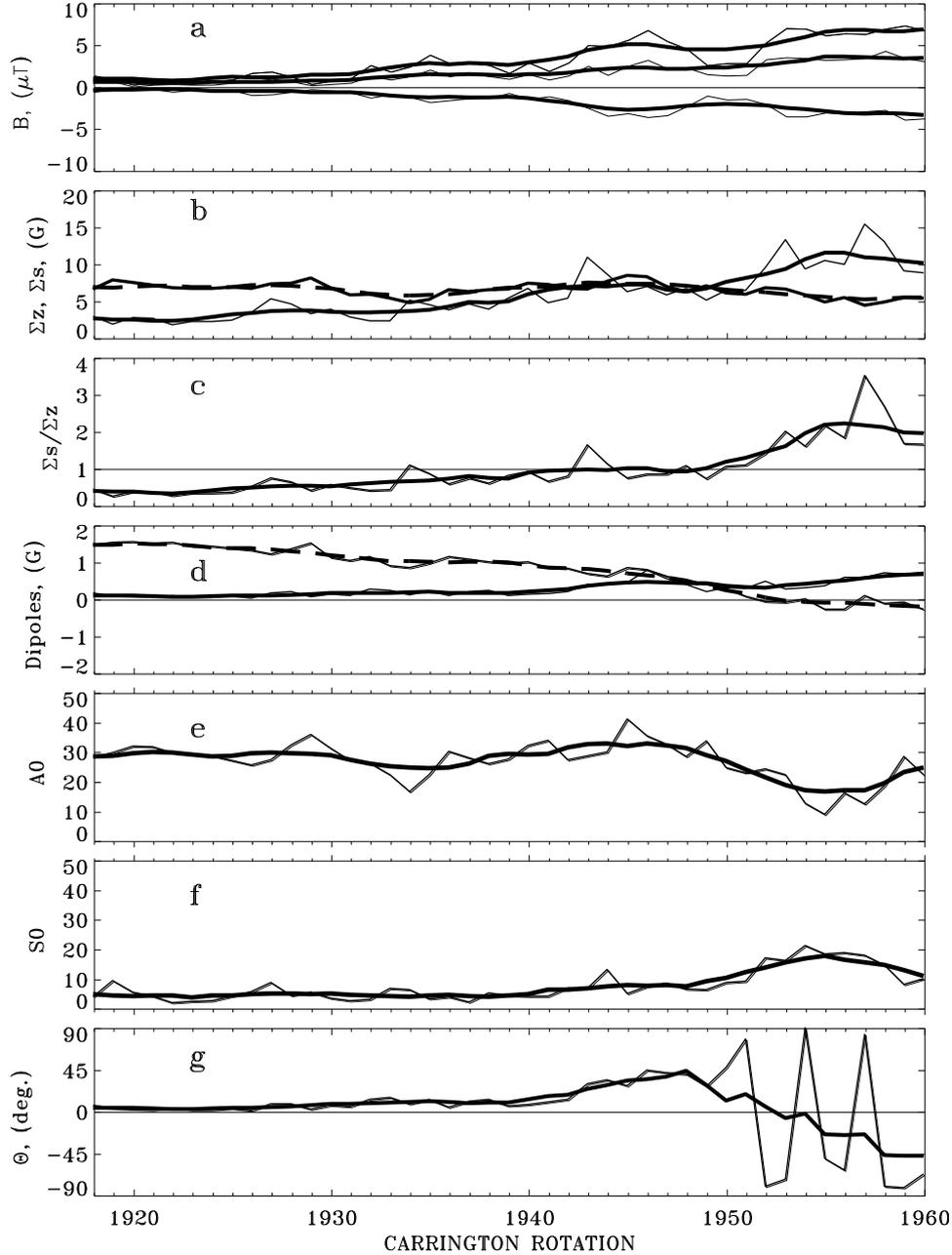

Figure 1. (a) Changes in the magnetic fields of positive- and negative-polarity magnetic fields and the sum of their modules on the source surface (2.5 Rs), calculated in the potential approximation; (b) The sum of zonal (dashed line) and the sum of sectorial harmonics (solid line); (c) The ratio of the sum of the sectorial to zonal harmonics; (d) the axisymmetric components of the dipole (dashed line) and equatorial dipole (solid line); (e) Axisymmetric and antisymmetric relative to the equator harmonic spectrum; (f) Axisymmetric and symmetric with respect to the equator harmonics; (g) Changes in the polar angle of the dipole component. The thin lines correspond to the data for each CR, and thick lines correspond to the average for 7 CRs.

## 4. EVOLUTIONARY CHANGES OF CHs DURING THE RISING PHASE OF CYCLE 23

CHs are known to be good tracers of cyclic variations of solar GMF (Ikhsanov, Ivanov, 1999; Mogilevsky, Obridko, Shilova, 1997; Obridko, Shelting, 1989; Bilenko, 2002; Bilenko and Tavastsherna, 2016; Bilenko, Tavastsherna, 2017). During the solar activity minimum when the zonal structure of the GMF was dominated, the CHs followed GMF structure. Figure 2 shows the examples of the structure of GMF, as well as the distribution of photospheric magnetic fields and CHs according to daily data during the domination of the zonal structure. Figures 1a5, b5, c5, d5 demonstrate distribution of GMF calculated in the potential approximation on the source surface (2.5 Rs) for CRs 1919, 1922, 1924, and 1929. The images of the full disk of the Sun in the line 195 Å according to SOHO/EIT (a1, A2, b1, b2, c1, c2, d1, d2) and combined images of daily data of photospheric magnetic fields and CHs in the line HeI 10830 Å from the Kitt Peak observatory (a3, a4, b3, b4, c3, c4, d3, d4) for different longitudinal intervals are shown above. The arrows on the magnetic field maps indicate the dates of the daily CH images for the left and right images respectively. It is seen that during the zonal structure domination the large CHs are observed at the poles. Often they have large extensions to low latitudes. Sometimes these extensions are disconnected from the polar CHs and form separate CHs in the same hemisphere in the regions of photospheric magnetic fields of the same polarity as the magnetic fields of the corresponding pole of the Sun. It is known that almost all high-latitude CHs are genetically related to the corresponding polar CHs (Bohlin, 1977). At low latitudes, separate small short-lived CHs are formed in the regions of magnetic fields with the polarity of the polarity of the magnetic field at the corresponding pole. The zonal structure of GMF and the zonal distribution of CHs, despite short-term disturbances, in general, remained stable during its dominance in the cycle minimum up to CR 1941. Since the beginning of the growth of sectorial harmonics (CR 1926), short-term disturbances of the neutral magnetic field line on the source surface have been observed.

With the growth of activity and the growth of sectorial harmonics of GMF (CR 1941) the distribution of CHs was also changed. Dynamics of the latitudinal distribution of non-polar CHs traced changes in the structure of the GMF from the zonal to the sectorial. The CHs of one polarity occupied the regions extended in latitude and longitude (Bilenko, 2002; Bilenko and Tavastsherna, 2016). Figure 3 shows some examples of different GMF topologies on the source surface at different states of the sectorial structure for CRs 1943, 1948, 1951, and 1958 (Figure 3, a5, b5, c5, d5) and daily CH images in EUV 195 Å (a1, a2, b1, b2, c1, c2, d1, d2) and line HeI 10830 Å, combined with photospheric magnetic fields (a3, a4, b3, b4, c3, c4, d3, d4). It is seen that the polar CHs decreasd sharply. They disappeared to the maximum of the cycle and reappeared after the change of the sign of the GMF. Non-polar CHs, corresponding to the photospheric magnetic fields of one po larity, began to form extended latitudinal-longitudinal CHs, reflecting the distribution of GMF. The location of the CHs in the regions of photospheric magnetic fields of positive or negative polarity on the solar disk corresponded to the topology of the GMF in the corresponding CR. Thus, CHs corresponding to positive photospheric magnetic fields are located in the zones of extended latitude and longitude of GMF structures of the same polarity.

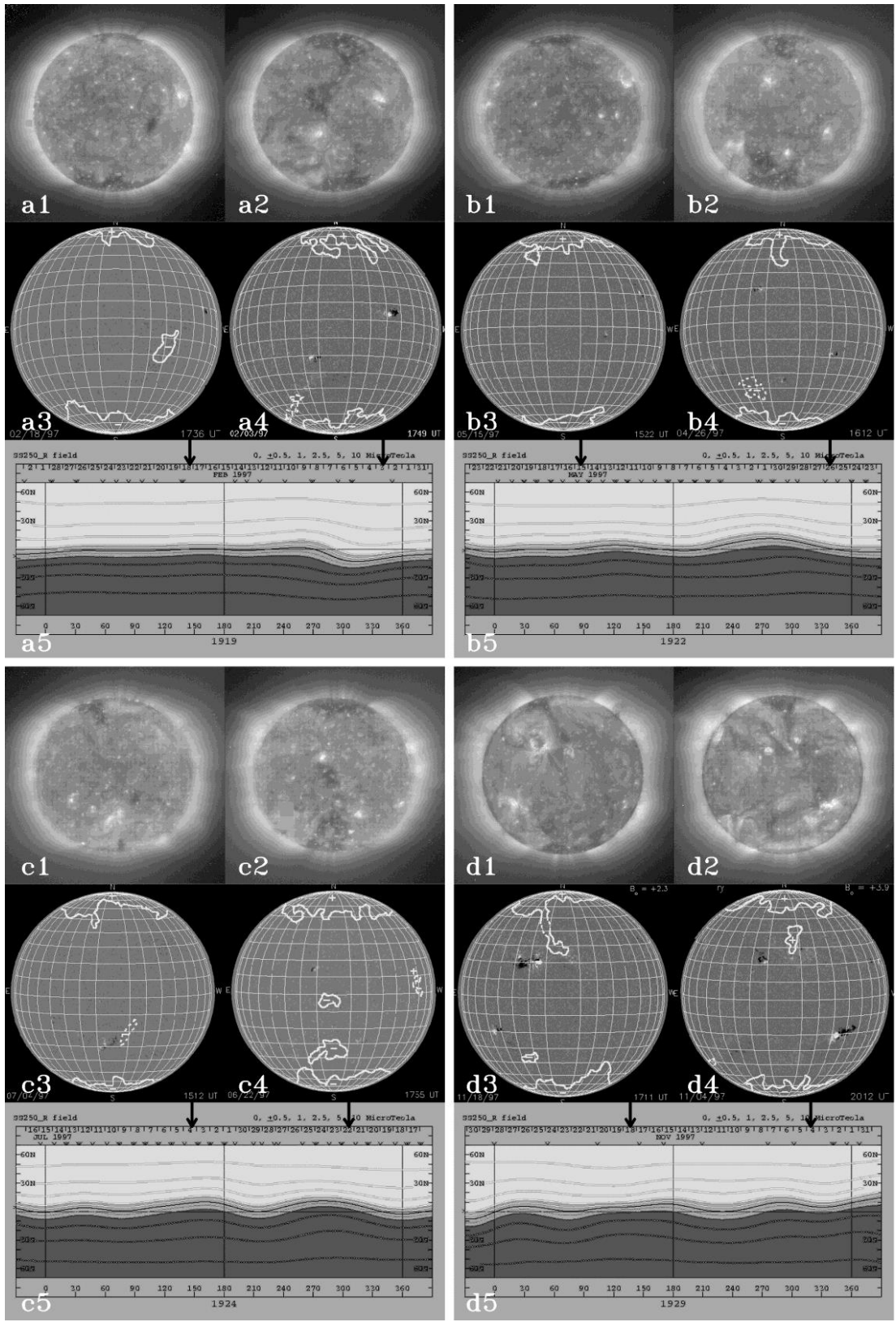

Figure 2. Examples of zonal structure of GMF and distribution of daily photospheric magnetic fields and CHs. GMF calculated in potential approximation on the source surface (2.5 Rs) for CRs 1919, 1922, 1924, and 1929 (a5, b5, c5, d5), images of the full disk of the Sun in the line 195 Å according to SOHO/EIT (a1, A2, b1, b2, c1, c2, d1, d2) and combined images of daily data of photospheric magnetic fields and CHs in the line HeI 10830 Å of the Kitt Peak observatory (a3, a4, b3, b4, c3, c4, d3, d4). The arrows indicate the dates of the daily CH images for the left and right images, respectively.

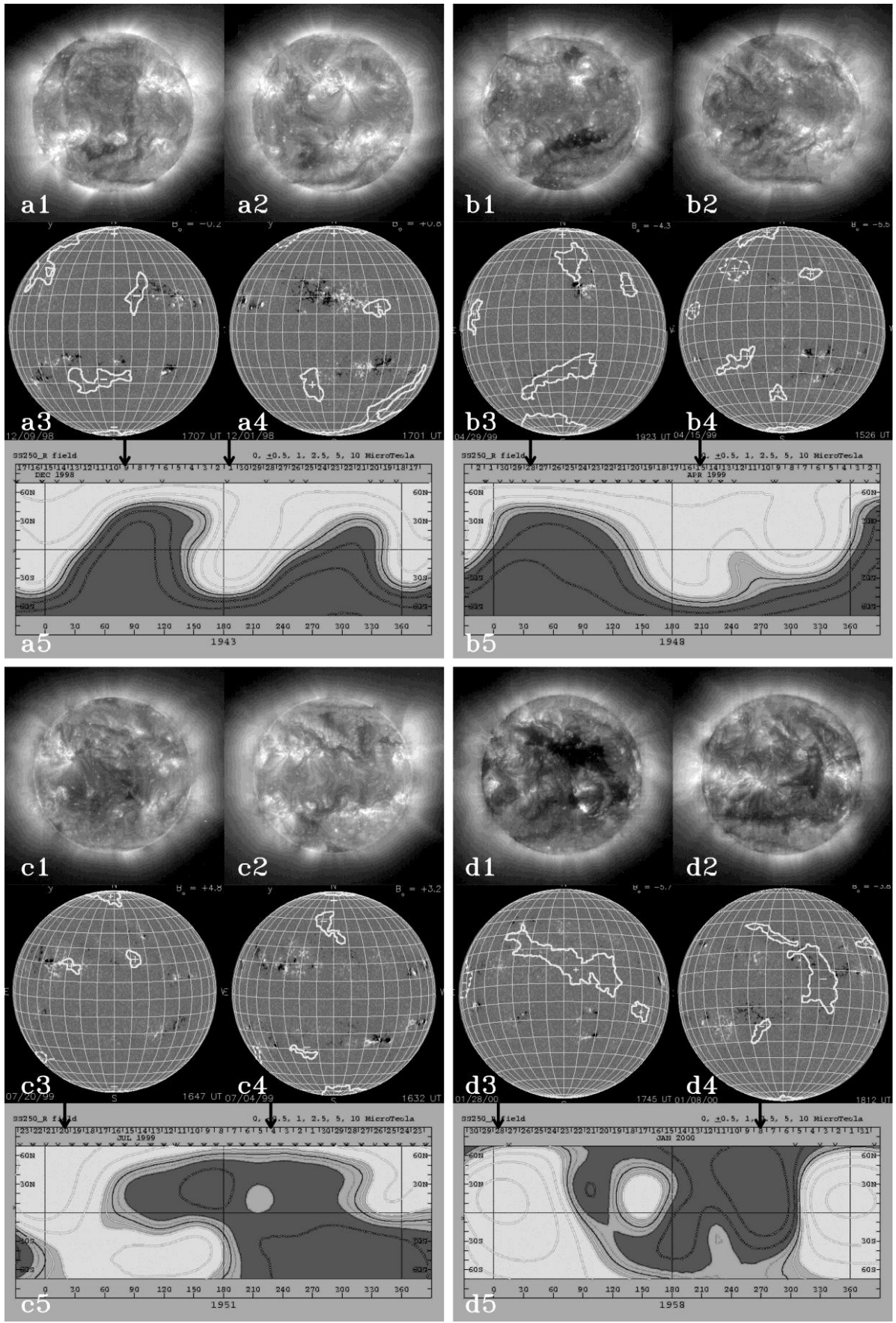

Figure 3. Examples of the sectorial structure of GMF for CRs 1943, 1948. 1951, and 1958. The arrangement of the images and their designations are identical to Figure 2.

In contrast to the zonal structure, the sectorial structure of the GMF was not stable, but underwent sharp changes, which were reflected in the latitudinal-longitudinal redistribution of positive- and negative-polarity magnetic fields. During the rising phase of cycle 23, the transformations of the structure from four-sector to two-sector and inversely were observed. Figures 4-6 show the synoptic maps of GMF, on the right, and combined maps of photospheric magnetic fields and CHs, on the left, for successive CRs during the period of domination of the sectorial structure of GMF. From CR 1932 a clear four-sector structure of the GMF was observed (Figure 4). At the longitudes of about 60° to 180° and 330°-360° magnetic fields of negative-polarity, and at the longitudes 0°-60° and 180°-330° the positive-polarity magnetic fields were dominated. This structure of the GMF, undergoing minor changes, remained approximately the same in CR 1933. In CRs 1934-1936 the zone of domination of positive-polarity magnetic fields sharply shifted from longitudes 180°-330° to longitudes 120°-240°, but, in general, the topology of GMF remained the same. The location of the CHs clearly corresponded to the distribution of GMF and followed its changes. Attention should be paid to the formation of CHs in the region of positive GMF (~180° longitude and ~40° latitude, southern hemisphere). The latitudinal redistribution of positive and negative magnetic fields began from CR 1937. In 1937, the structure of GMF for a short time became two-sector. In CRs 1938-1949 there were quite significant changes in the topology of the magnetic field, although the overall structure of the GMF remained four-sector (Figure 4, 5). From CR 1940 to CR 1943 there was a fairly stable four-sector structure with the dominance of negative-polarity magnetic fields at the longitudes of about 45°-150° and 270°-340°, and positive-polarity magnetic fields at the longitudes of about 0°-45°, 150°-270° and 340°-360°. CHs were also formed in the regions of negative-polarity GMF only in the regions of negative photospheric magnetic fields, and in the regions of positive-polarity GMF only in the regions of positive photospheric magnetic fields. In CR 1944 the structure dramatically transformed into the two-sector. Figure 5 shows the period from CR 1943 to CR 1951 when this reorganization of the GMF occurred. On the longitudes of 30°-180° the of negative-polarity and in the longitudes 180°-360° the positive-polarity magnetic fields were dominated. During CRs 1945-1948, this structure became more pronounced, and from CR 1949 the shift of the zone of domination of positive magnetic fields to the decreasing longitude continuing until CR 1952, when the distribution of magnetic fields changed to the opposite with respect to that existing during the CRs 1944-1948. At the same time, the structure remained two-sector all the time. The change in the distribution of CHs corresponding to positive- and negative-polarity photospheric magnetic fields followed the dynamics of GMF. In CR 1947 at the longitudes 0°-140° negative-polarity GMF were dominated and CHs formed in the regions of negative-polarity photospheric magnetic fields, and at the longitudes 150°-360° positive-polarity photospheric magnetic fields were dominated and CHs formed in the regions of positive photospheric magnetic fields. In CRs 1948-1950, the GMF was reorganized, which was accompanied by an increase in the contribution of the axisymmetric and symmetric relative to equator components and a decrease in the axisymmetric but antisymmetric components relative to the equator, and by CR 1951 the distribution of the magnetic fields of the GMF changed to the opposite. CHs were also formed at longitudes 0°-150° in the regions of positive-polarity photospheric magnetic fields to CR 1951, and at longitudes 160°-360° on the regions of negative-polarity photospheric magnetic fields. In CR 1953, the structure of GMF and the distribution of CHs again sharply changed and a stable four-sector structure was formed, existing up to CR 1955 inclusive with the distribution of positive-polarity magnetic fields at longitudes from about 80° to 180°, and negative-polarity ones at longitudes 0°-80°, 180°-240° and 315°-360° (Figure 6). Since CR 1956, the GMF structure has been transformed into a two-sector one with the formation of separate closed structures of one polarity, which is typical for the period of solar activity maximum, when a single neutral magnetic field line, which is the base of the heliospheric current sheet, forms a complex topology within the two-sector structure of the GMF. As example, this can be seen in CRs 1958, 1959 (Figures 3, 6). It should be noted that during the periods of sharp reorganization of GMF the area of CHs decreases, or CHs are not formed at all.

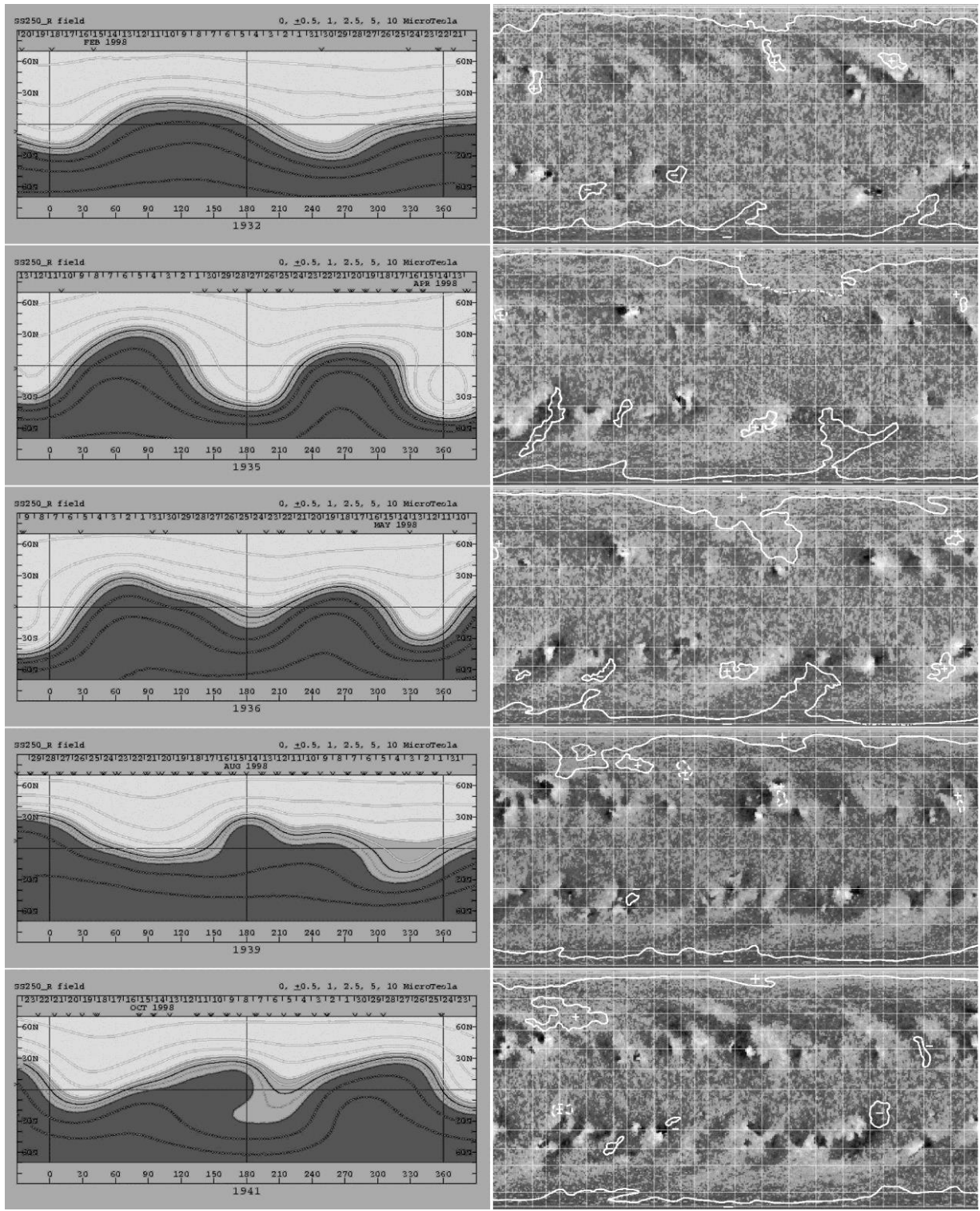

Figure 4. Dynamics of the sectorial structure of GMF, CRs 1932-1941. Left – distribution of GMF on the source surface, right – CHs superimposed on the photospheric magnetic fields. Light tones denote positive-polarity magnetic fields, while dark tones denote negative-polarity ones.

Comparison of changes in the zones of formation of CHs with the dynamics of local magnetic fields of active regions (AOs) showed that for a number of CHs observed during this period, the changes were not associated with the appearance or dissipation of specific AO. Thus, the appearance of a number of high-latitude CHs in the regions of photospheric magnetic fields of opposite polarity to the sign of the field of the corresponding pole, for example in CRs 1935, 1936, 1945, 1946, were not associated with AOs. With the growth of activity, the number of AOs increased greatly and it is difficult to separate the evolutionary changes in CHs and AOs. It is important to note that regardless of the dynamics of the local fields of AOs, CHs are formed in the regions of one polarity of the photospheric magnetic fields corresponding to the unipolar zones of GMF at large latitudinal-longitudinal intervals, although in these unipolar zones of GMF there are large areas of photospheric magnetic fields of opposite polarity, but where CHs are not formed during the lifetime of these regions. From the study it follows that the formation of CHs are determined by the evolution of the GMF.

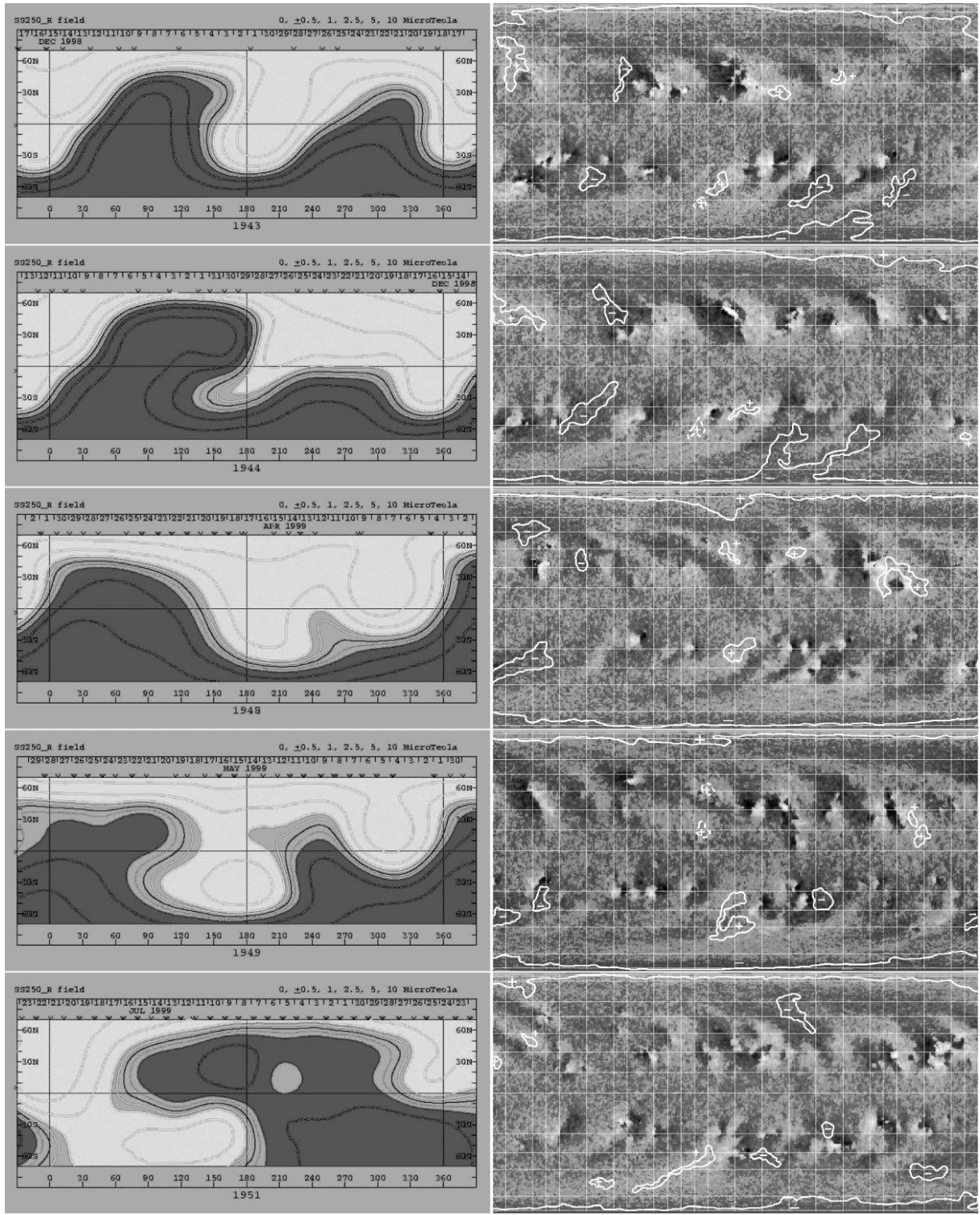

Figure 5. Dynamics of the sectorial structure of GMF, CRs 1943-1951. The description of the images is identical to Figure 4.

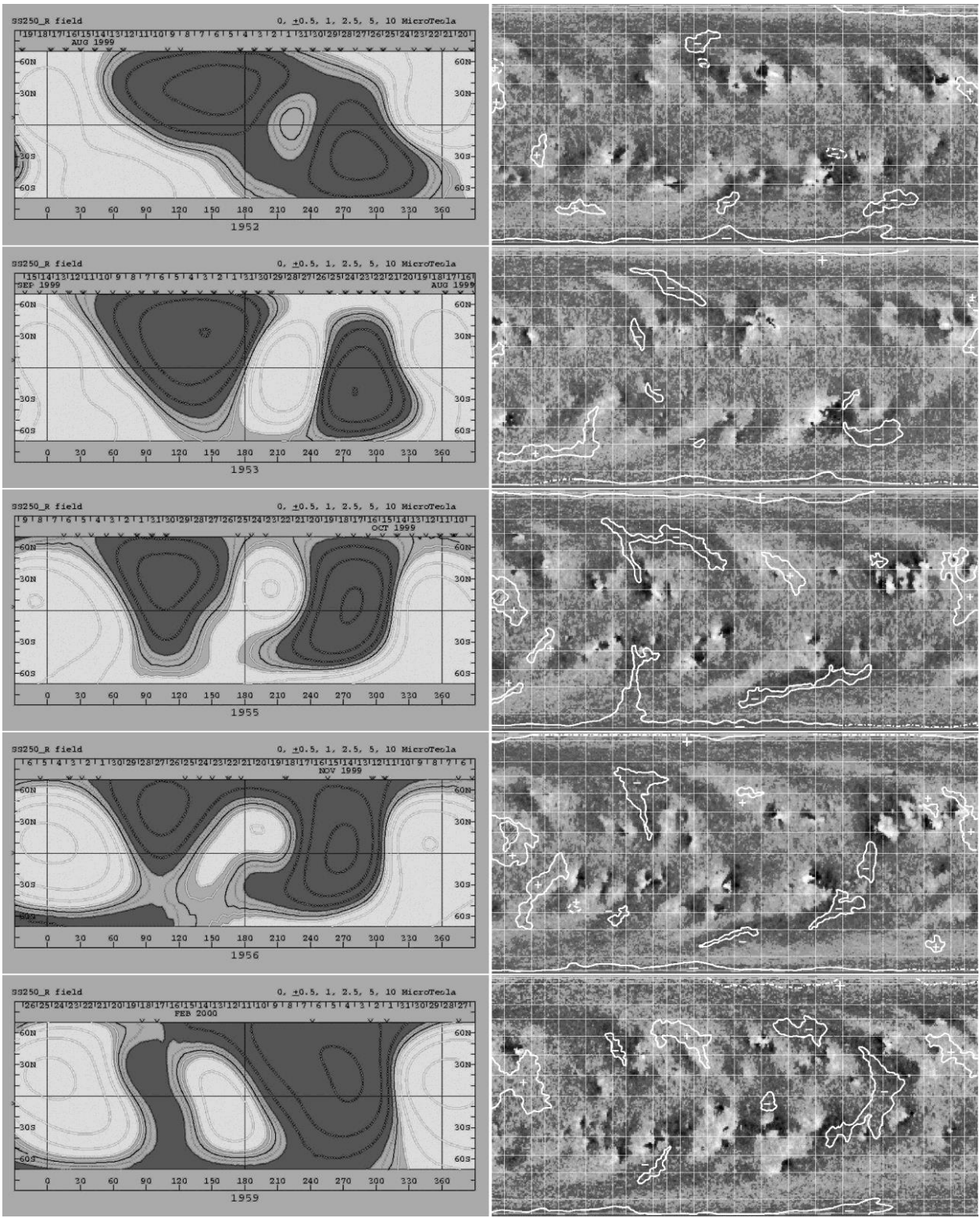

Figure 6. Dynamics of sectorial structure of GMF, CRs 1952-1958. The description of the images is identical to Figure 4.

## 5. SUMMARY

The paper considers the dependence of the dynamics of CHs on the evolutionary changes in the structure of GMF at the rising phase of cycle 23 from 01.01 1997 to 01.03.2000 (CRs 1918-2059).

It was shown that at the rising phase of solar activity in cycle 23 there was a transition from the zonal to the sectorial structure of GMF. The zonal structure was quasi-stable. The sum of zonal harmonics dominated up to CR 1941, although a stable four-sector structure of GMF was formed from 1932. In CRs 1941-1950 the contribution of the zonal and sectoral component became approximately equal, and from CR 1950 the sectorial structure of GMF was dominated.

Sectorial structure of GMF underwent sharp a change from four-sector in the early growth of the sectorial harmonics to the two-sector, then back to four-sector and then again to the two-sector. CHs unambiguously traced all evolutionary changes of GMF and its reorganizations from the zonal structure to the sectorial and further topological changes in the sector structure of GMF. Longitude-latitude distribution of CHs corresponded directly to the longitude-latitude distribution of the GMF. While large areas of photospheric magnetic fields located near the CHs, that were in the regions of photospheric magnetic fields with a polarity corresponding to the sign of the GMF, may have the polarity opposite to the polarity of the GMF for this zone, the CHs were not formed on them.

During the reorganization of the structure of GMF, the area of CHs either decreased or they were not formed at all.

Since the evolution of GMF and, accordingly, CHs, are cyclical, the obtained results may be of interest for the purposes of the space weather and geomagnetic activity forecast.

## ACKNOWLEDGEMENTS

Wilcox Solar Observatory data used in this study were obtained via the web site http://wso.stanford.edu at 2018:03:11 01:13:34 PST courtesy of J.T. Hoeksema. The Wilcox Solar Observatory is currently supported by NASA.

NSO/Kitt Peak data used here were produced cooperatively by NSF/NOAO, NASA/GSFC, and NOAA/SELL.

SOHO/EIT data were used. SOHO is a project of international cooperation between ESA and NASA.